\documentclass[12pt,a4paper]{article}
\usepackage{amsmath}
\usepackage[dvips]{graphicx}
\input epsf
\usepackage{times}
\begin{document}
\begin{titlepage}
\title{Chiral quark filtering mechanism of hyperon polarization}
\author{S. M. Troshin,
 N. E. Tyurin\\[1ex]
\small  \it Institute for High Energy Physics,\\
\small  \it Protvino, Moscow Region, 142281, Russia} \normalsize
\date{}
\maketitle

\begin{abstract}
The model combined with unitarity and impact parameter picture  provides a rather
simple mechanism for generation of hyperon polarization  in collision of unpolarized hadrons.
We  concentrate  on a particular problem of $\Lambda$-hyperon polarization and derive
its linear $x_F$-dependence as well as its energy and transverse momentum
independence at large $p_\perp$ values. Mechanism
 responsible for the single--spin asymmetries  in pion production is also discussed.

\end{abstract}
\end{titlepage}
\setcounter{page}{2}

\section*{Introduction}
One of the most interesting and persistent for a long time spin phenomena
was observed in inclusive hyperon production in collisions of
unpolarized hadron beams. A very significant polarization of
$\Lambda$--hyperons has been discovered almost three  decades ago \cite{newrev}.
 There is a
list of theoretical models which relate polarization
mechanism with various aspects of hadron interaction dynamics \cite{felix} but
till now it has not obtained a satisfactory explanation. Experimentally the process
of $\Lambda$-production has been studied more extensively than other hyperon
production processes.
Therefore we will emphasize on the particular riddle  of $\Lambda$--polarization
because  spin structure of this particle is most simple and
 is determined by strange quark only. We also
 provide comments on how this mechanism can
be used for the explanation of single-spin asymmetries in the inclusive pion production.

It should be noted that understanding of
transverse single-spin asymmetries in DIS (in contrast to the hyperon polarization)
has observed significant progress during last years; this progress is related
to an account of final-state interactions from gluon exchange \cite{brodsky,metz} --
coherent effect not suppressed in the Bjorken limit.

Experimental
 situation with hyperon polarization is widely known and stable for a long time.
Polarization of $\Lambda$ produced in the unpolarized inclusive $pp$--interactions
is negative and energy
independent. It increases linearly with $x_F$ at large transverse momenta
($p_\perp\geq 1$ GeV/c),
and for such
values of transverse momenta   is almost
$p_\perp$-independent \cite{newrev}.

On the theoretical side,  perturbative QCD
with a straightforward collinear factorization scheme
leads to small values of $\Lambda$--polarization
\cite{pump,gold} which are far below of the corresponding experimental data.
Modifications of this scheme and  account for higher twists contributions allows
to obtain higher magnitudes of polarization but do not change
a decreasing  dependence proportional to
$1/p_\perp$ \cite{efrem,sterm,koike} at large transverse momenta.
It is difficult to reconcile this behavior  with the flat  experimental
data dependence on the transverse momenta. Inclusion of the internal transverse momentum
of partons ($k_\perp$--effects) into the
so called polarizing fragmentation functions  leads to similar decreasing polarization
 \cite{anselm}.
In addition it should be noted that the perturbative QCD has also problems in the description
of the unpolarized scattering, e.g. in inclusive cross-section
for $\pi^0$-production, at the energies lower than the RHIC energies \cite{bsof}.

The essential point  of the approaches mentioned above is that the vacuum at short distances
is taken to be a perturbative one.
There is an another possibility. It might happen  that
the hyperon polarization dynamics originates from the genuine
nonperturbative sector of QCD (cf. e.g. \cite{spin02}).
The point of view that the polarization has its roots hidden in the nonperturbative
sector of QCD is not an isolated one  and several approaches based on
 nonperturbative dynamics has appeared up to now.
We briefly mention them later.

In the nonperturbative sector of QCD the  two important
phenomena,  confinement and spontaneous breaking of chiral symmetry ($\chi$SB)\cite{mnh}
should be reproduced.
The  relevant scales   are characterized by the
parameters $\Lambda _{QCD}$ and $\Lambda _\chi $, respectively.  Chiral $SU(3)_L\times
SU(3)_R$ symmetry is spontaneously  broken  at the distances
in  the range between
these two scales.  The $\chi$SB mechanism leads
to generation of quark masses and appearance of quark condensates. It describes
transition of current into  constituent quarks.
  Constituent quarks are the quasiparticles, i.e. they
are a coherent superposition of bare  quarks, their masses
have a magnitude comparable to  a hadron mass scale.  Therefore
hadron  is often represented as a loosely bounded system of the
constituent quarks.
These observations on the hadron structure lead
to  understanding of several regularities observed in hadron
interactions at large distances. It is well known  that such picture  provides
reasonable  values  for the static characteristics of hadrons, for
 instance, their magnetic moments. The other well known direct result
   is  appearance of the Goldstone bosons.

The most recent  approach  to
single--spin asymmetries (SSA) based on nonperturbative QCD has been developed in \cite{burk} where, in particular,
$\Lambda$-polarization has been related to the large magnitude of the transverse
 flavor dipole moment of the transversely polarized baryons in the infinite momentum frame.
  It is based
on the parton picture in the impact parameter space and assumed specific helicity--flip
generalized parton distribution.

The instanton--induced mechanism of SSA generation was considered in \cite{koch,shur}
and relates those asymmetries with a genuine nonperturbative QCD interaction.
It should be noted that the physics of instantons (cf. e.g. \cite{inst})
can provide microscopic explanation for the $\chi$SB mechanism\footnote{We are
grateful to Dmitri Diakonov for the interesting communication on this
matter regarding the polarization phenomena.}.

We discuss here mechanism for hyperon polarization based on chiral
quark model\footnote{It  has been successfully applied
for the  explanation of the nucleon spin structure \cite{cheng}.} \cite{mnh} and the
filtering spin states related to unitarity in the $s$-channel.
This mechanism connects polarization with  asymmetry in the
position (impact parameter) space.

\section{Chiral quark model and the mechanism of spin states filtering}
As it was already mentioned constituent quarks and Goldstone bosons are the effective
degrees of freedom in the chiral quark model. We consider a
 hadron consisting of the valence
constituent quarks located in the central core which is embedded into  a quark
condensate. Collective excitations of the condensate are the Goldstone bosons
and the constituent quarks interact via exchange
of Goldstone bosons; this interaction is mainly due to a pion field which is of the flavor--
 and spin--exchange nature. Thus, quarks generate a strong field which
binds them \cite{diak}.

At the first stage of hadron interaction common effective
self-consistent field is appeared.
Valence constituent quarks   are
 scattered simultaneously (due to strong coupling with Goldstone bosons)
and in a quasi-independent way by this effective strong
 field. Such ideas were already used in the model \cite{csn} which has
been applied to description of elastic scattering and hadron production \cite{mult}.

The initial state  particles (protons) are unpolarized.
It means that states with spin up and spin down have equal probabilities.
The main idea of the proposed mechanism is the  filtering
of the two initial spin states of equal probability due to different strength of interactions. The particular
mechanism of such filtering can be developed on the basis of chiral quark model,
formulas for inclusive cross section (with account for the unitarity) \cite{tmf} and
notion on the quasi-independent nature of valence quark scattering in the effective field.

We will exploit the feature of chiral quark model that constituent quark $Q_\uparrow$
with transverse spin in up-direction can fluctuate into Goldstone boson and
  another constituent quark $Q'_\downarrow$ with opposite spin direction,
   i. e. perform a spin-flip transition \cite{cheng}:
\begin{equation}\label{trans}
Q_\uparrow\to GB+Q'_\downarrow\to Q+\bar Q'+Q'_\downarrow.
\end{equation}
An absence of arrows means that the corresponding quark is unpolarized.
To compensate quark spin flip $\delta {\bf S}$ an orbital angular momentum
$\delta {\bf L}=-\delta {\bf S}$ should be generated in final state of reaction (\ref{trans}).
The presence of this orbital momentum $\delta {\bf L}$  in its turn
means  shift in the impact parameter
value of the final quark $Q'_\downarrow$ (which is transmitted to the shift in the impact
parameter of $\Lambda$)
\[
\delta {\bf S}\Rightarrow\delta {\bf L}\Rightarrow\delta\tilde{\bf b}.
\]
Due to   different strengths of interaction at the different values of the
impact parameter, the processes of transition to the
spin up and down states will have different probabilities which  leads eventually to
polarization of $\Lambda$.

\begin{figure}[h]
\begin{center}
  \resizebox{4cm}{!}{\includegraphics*{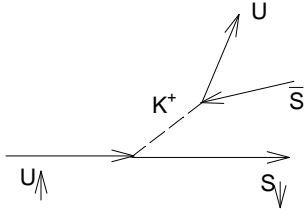}}
\end{center}
\caption{Transition of the spin-up constituent quark $U$ to the spin-down strange quark.
 \label{ts1}}
\end{figure}
In a particular case of $\Lambda$--polarization the relevant transitions
of constituent quark $U$ (cf. Fig. 1) will be correlated with the shifts $\delta\tilde b$
in impact parameter $\tilde b$ of the final
$\Lambda$-hyperon, i.e.:
\begin{eqnarray}
  \nonumber U_\uparrow & \to & K^+ + S_\downarrow\Rightarrow\;\;-\delta\tilde{\bf b} \\
\label{spinflip} U_\downarrow & \to & K^+ + S_\uparrow\Rightarrow\;\;+\delta\tilde {\bf b}.
\end{eqnarray}
Eqs. (\ref{spinflip}) clarify mechanism of the filtering of spin states:
 when shift in impact
parameter is $-\delta\tilde {\bf b}$ the
interaction is stronger compared to the case when shift is $+\delta\tilde {\bf b}$,
and the final $S$-quark
(and $\Lambda$-hyperon) is polarized negatively.

It is important to note here that the shift of $\tilde{\bf b}$
(the impact parameter of final hyperon)
is translated to the shift of the impact parameter of the initial particles according
to the relation between impact parameters in the multiparticle production\cite{webb}:
\begin{equation}\label{bi}
{\bf b}=\sum_i x_i{ \tilde{\bf  b}_i}.
\end{equation}
The variable $\tilde b$ is conjugated to the transverse momentum of $\Lambda$,
but relations  between functions depending on the impact parameters
$\tilde b_i$,  which will be used further for the calculation of polarization,
are nonlinear and therefore we
will use semiclassical correspondence between small and large values
 of transverse momentum and impact parameter:
\begin{equation}\label{bp}
\mbox{small}\;\tilde b \Leftrightarrow \mbox{large}\;p_\perp,
\end{equation}
\begin{equation}\label{pb}
\mbox{large}\;\tilde b \Leftrightarrow \mbox{small}\;p_\perp.
\end{equation}

We consider production of $\Lambda$ in the fragmentation region, i.e.
at large $x_F$ and therefore use approximate relation
\begin{equation}\label{bx}
b\simeq x_F\tilde b,
\end{equation}
which results from Eq. (\ref{bi})\footnote{We make here an additional assumption on the
small values of Feynman $x$ for other particles}.

The mechanism of the polarization generation  is quite natural
 and has an optical analogy with the passing
 of the unpolarized light through the glass of polaroid.
The particular mechanism of filtering of spin states is related to the
 emission of Goldstone bosons by constituent quarks. This picture is more physically
  apparent and justified than the polarization generation due to multiple
   scattering of constituent quarks in the effective field developed
earlier by the authors in \cite{uspoll}.

We will  now obtain an expression for the polarization which takes into account
unitarity in the direct channel of reaction and apply chiral quark filtering to conclude
on polarization dependence on the kinematical variables.

\section{Polarization dependence on kinematical variables}
We use the explicit formulas for inclusive
cross--sections of the process
\[ h_1 +h_2\rightarrow h_3^\uparrow +X, \] where hadron $h_3$ is a hyperon whose
transverse polarization is measured, obtained in
\cite{tmf}. The main feature of this formalism is an account of
unitarity in the direct channel of
reaction. The corresponding formulas have the form
\begin{equation}
{d\sigma^{\uparrow,\downarrow}}/{d\xi}= 8\pi\int_0^\infty
bdb{I^{\uparrow,\downarrow}(s,b,\xi)}/ {|1-iU(s,b)|^2},\label{un}
\end{equation}
where $b$ is the impact  parameter of the initial
particles. Here the function
$U(s,b)$ is the generalized reaction matrix (for unpolarized scattering)
which is determined by the basic dynamics of elastic scattering.
 The elastic scattering amplitude in the impact
parameter representation $F(s,b)$
   is related \cite{log,fey,cudell} then
  to the function $U(s,b)$ by the  relation:
  \begin{equation} F(s,b)=U(s,b)/[1-iU(s,b)].
\label{6} \end{equation}
This relation allows one to obey unitarity provided inequality
  $ \mbox{Im}\,U(s,b)\geq 0\,$  is fulfilled.

The functions $I^{\uparrow,\downarrow}$ in Eq. (\ref{un}) are related   to the
functions  $|U_n|^2$, where $U_n$  are the multiparticle
analogs of the $U$ \cite{tmf}. The kinematical variables $\xi$
($x_F$ and $p_\perp$) describe the state of the produced particle
$h_3$.
   Arrows $\uparrow$ and $\downarrow$ denote
   transverse spin directions of the final hyperon $h_3$.

Polarization  \[ P=
\{\frac{d\sigma^\uparrow}{d\xi}-\frac{d\sigma^\downarrow}{d\xi}\}/
\{\frac{d\sigma^\uparrow}{d\xi}+\frac{d\sigma^\downarrow}{d\xi}\} \]
can be expressed in terms of the functions $I_{-}$, $I_{0}$ and $U$:
\begin{equation} P(s,\xi)=\frac{\int_0^\infty bdb
I_-(s,b,\xi)/|1-iU(s,b)|^2} {2\int_0^\infty bdb
I_0(s,b,\xi)/|1-iU(s,b)|^2},\label{xnn}
\end{equation}
where $I_0=1/2(I^\uparrow+I^\downarrow)$ and $I_-=(I^\uparrow-I^\downarrow)$.

Now we turn to the functions $I^\uparrow $ and $I^\downarrow $
and will apply chiral
quark model to get an information on the hyperon polarization
and its dependence on energy, $x_F$ and $p_\perp$.
On the basis of the described chiral quark filtering mechanism we can
assume that the functions
$I^\uparrow(s,b,\xi)$ and $I^\downarrow(s,b,\xi)$ are related to the functions
$I_0(s,b,\xi)|_{\tilde b+\delta\tilde b }$ and $I_0(s,b,\xi)|_{\tilde b-
\delta\tilde {b} }$,
respectively, i.e.
\begin{equation}\label{der}
I_-(s,b,\xi)=I_0(s,b,\xi)|_{\tilde {b}+\delta\tilde {b} }-
I_0(s,b,\xi)|_{\tilde{b}-\delta\tilde{b} }
=2\frac{\delta I_0(s,b,\xi)}{\delta\tilde{b}}\delta\tilde b.
\end{equation}

We can connect $\delta\tilde b$ with the radius of quark interaction
$r_{U\to S}^{flip}$
responsible for the transition $U_\uparrow\to S_\downarrow$ changing quark spin and flavor:
\[
\delta\tilde b\simeq r_{U\to S}^{flip}.
\]

Using the above formulas and,
in particular, relation (\ref{bx}), we can write
the following expression for polarization $P_\Lambda(s,\xi)$
\begin{equation} P_\Lambda(s,\xi)\simeq x_Fr_{U\to S}^{flip}\frac{\int_0^\infty bdb
I'_0(s,b,\xi)db/|1-iU(s,b)|^2} {\int_0^\infty bdb
I_0(s,b,\xi)/|1-iU(s,b)|^2},\label{poll}
\end{equation}
where $I'_0(s,b,\xi)={dI_0(s,b,\xi)}/{db}$. We have made
replacement in (\ref{poll}) according to relation (\ref{bx}):
\[
{\delta I_0(s,b,\xi)}/{\delta\tilde{b}}\to {dI_0(s,b,\xi)}/{db}.
\]

It is clear that
polarization of $\Lambda$ - hyperon (\ref{poll})
should be negative because $I'_0(s,b,\xi)<0$.

The function $U(s,b)$ is
chosen  as a product of the averaged quark amplitudes
in accordance with the quasi-independence of valence constituent
quark scattering in the mean field \cite{csn}.

The generalized
reaction matrix $U(s,b)$ (in a pure imaginary case, which we consider
 here for simplicity) is
the following \footnote{This form leads, in particular, to asymptotic $\ln^2 s$ dependence
for total cross--sections and  $(1/s)^{N+3}f(\theta)$ dependence of differential
cross--sections at large angles \cite{csn}.}
\begin{equation} U(s,b) = i\tilde U(s,b)=ig(s)\exp(-Mb/\zeta ),
 \label{x}
\end{equation}
where
\[
g(s)\equiv g_0g^N_Q(s)\equiv g_0\left [1+\alpha
\frac{\sqrt{s}}{m_Q}\right]^N,
\]
$M$ is the total mass of $N$ constituent quarks with mass $m_Q$ in
the initial hadrons; $\alpha$ and $g_0$ are the parameters of
model. Parameter $\zeta$ is the one which determines a universal scale for
the quark interaction radius, i.e. $r_Q=\zeta /m_Q$.

Performing integration by parts we can rewrite
 the expression for polarization $P_\Lambda(s,\xi)$
in the form:
\begin{equation}\label{gpol}
P_\Lambda(s,\xi)\simeq -x_Fr_{U\to S}^{flip}
\frac{M}{\zeta}\frac{\int_0^\infty bdb
I_0(s,b,\xi)\tilde U(s,b) /[1+\tilde U(s,b)]^3} {\int_0^\infty bdb
I_0(s,b,\xi)/[1+\tilde U(s,b)]^2},
\end{equation}
To evaluate polarization dependence on $x_F$ and $p_\perp$
we use semiclassical correspondence  between transverse momentum and impact parameter
  values, i.e. (\ref{bp}) and (\ref{pb}).

Choosing the region of  small $p_\perp$ we select the large values of impact parameter
 and therefore  we  have
\begin{equation} P_\Lambda(s,\xi)\propto -x_Fr_{U\to S}^{flip}
\frac{M}{\zeta}\frac{\int_{b>R(s)} bdb
I_0(s,b,\xi)\tilde U(s,b) } {\int_{b>R(s)} bdb
I_0(s,b,\xi)},\label{pollsm}
\end{equation}
where $R(s)\propto \ln s$ is the hadron interaction radius, which serve as a scale
of large and small impact parameter values.
At large values of impact parameter $b$:
$\tilde U(s,b)\ll 1$ for $b\gg R(s)$ and therefore
 we will have small polarization $P_\Lambda\simeq 0$
in the region of small and moderate $p_\perp\leq 1$ GeV/c.

But at small values of $b$ (and large $p_\perp$): $\tilde U(s,b)\gg 1$
 and the following approximate relations
are valid
\begin{equation}
\int_{b<R(s)} bdb\frac{I_0(s,b,\xi)\tilde U(s,b)}{ [1+\tilde U(s,b)]^{3}}
\simeq
\int_{b<R(s)} bdb
I_0(s,b,\xi)\tilde U(s,b)^{-2}\label{simil},
\end{equation}
since we can neglect unity in the denominators of the integrands.
\begin{figure}[htb]
\begin{center}
  \resizebox{6cm}{!}{\includegraphics*{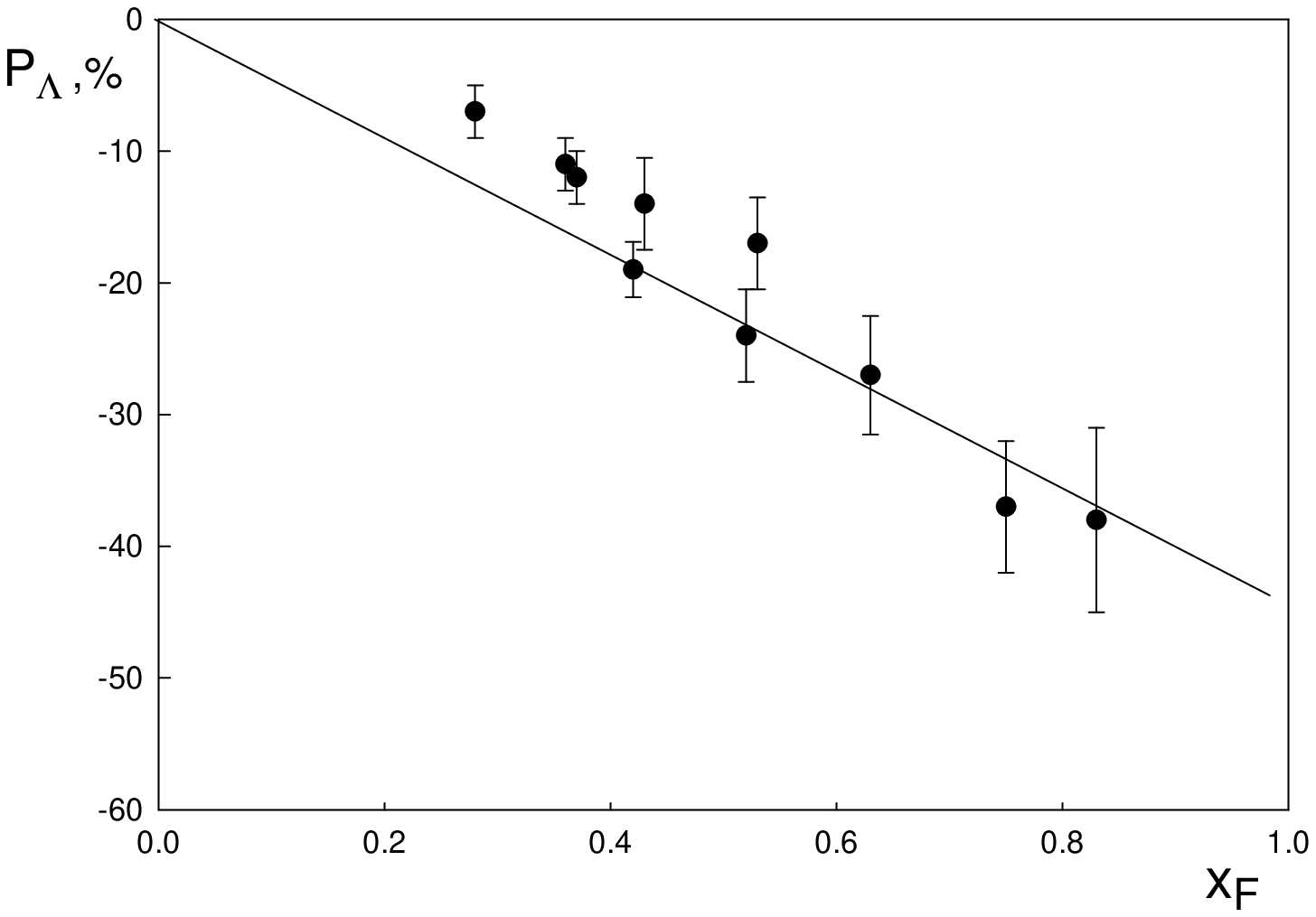}}\;\;\quad
  \resizebox{6cm}{!}{\includegraphics*{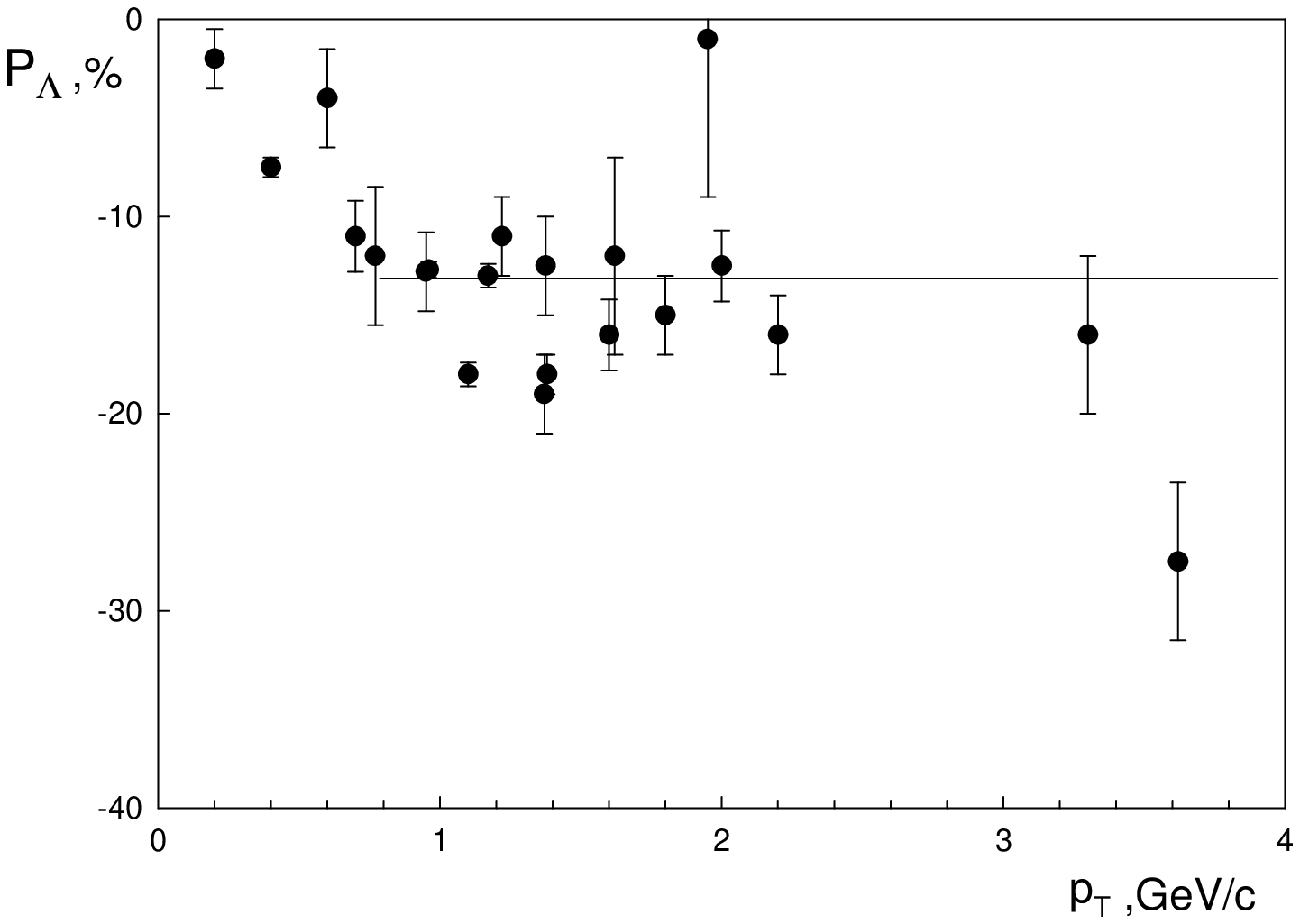}}
\end{center}
\caption{$x_F$ (left panel) and $p_T$ (right panel)
 dependencies of the $\Lambda$-hyperon
polarization} \label{ts}
\end{figure}
Thus, in (\ref{gpol}) the ratio of two integrals is of order
of unity and therefore  the energy and $p_\perp$-independent behavior
of polarization $P_\Lambda$ takes place at large values of $p_\perp$:
\begin{equation} P_\Lambda(s,\xi)\propto -x_Fr_{U\to S}^{flip}
{M}/\zeta.\label{polllg}
\end{equation}
This flat transverse momentum dependence results from the similar
rescattering effects for the different spin states, i.e. spin--flip and
non spin-flip interactions undergo similar absorption at short distances and
the relative magnitude of this absorption does not depend on energy. It is one
of the manifestations of unitarity.
The numeric value of polarization $P_\Lambda$ can be large: there are
no small factors in (\ref{polllg}). In (\ref{polllg}) $M$ is
proportional to two nucleon masses, the value of parameter $\zeta \simeq 2$.
 We expect that $r_{U\to S}^{flip}\simeq
0.1-0.2$ fm on the basis of the model \cite{csn,tmf}, however,
this is a crude estimate. The above qualitative
 features of polarization dependence on $x_F$,
$p_\perp$ and energy are in a good agreement with the experimentally observed trends
 \cite{newrev}.
For example, Fig. 2 demonstrates that the linear $x_F$ dependence is in a good agreement with
the experimental data in the fragmentation region ($x_F\geq 0.4$) where the model
should work. Of course,
the conclusion on the $p_\perp$--independence of polarization is a rather approximate one
and deviation from such behavior cannot be excluded.

\section{Inclusive cross-sections of the unpolarized $\Lambda$
production}
To demonstrate the model self-consistency we consider  in this section
the  unpolarized cross-section of $\Lambda$-production:
\begin{equation}
\frac{d\sigma}{d\xi}= 8\pi\int_0^\infty
bdb\frac{I_0(s,b,\xi)}{|1-iU(s,b)|^2}\label{unp}.
\end{equation}

In the fragmentation region we can simplify the problem and consider
the process of $\Lambda$-production as a quasi two-particle reaction,
where the second final particle has a mass $M^2\simeq (1-x_F)s$.
 The amplitude of this quasi two-particle reaction in the pure imaginary
 case (which we consider for simplicity) can be written in the form
 \begin{equation}
F(s,p_\perp,x_F)= \frac{is}{x_F^2\pi^2}\int_0^\infty
bdbJ_0(bp_\perp/x_F)\frac{I^{1/2}_0(s,b,x_F)}{1+U(s,b)}\label{amp}.
\end{equation}
To obtain Eq. \ref{amp} we used relations $b\simeq x_F\tilde{b}$ and due to
the fact that the functions $I_0$ is   quadratic on the the multiparticle
analog of the generalazed reaction matrix $U$, we use the relation
\begin{equation}\label{img}
I_0^{1/2}(s,b,p_\perp, x_F)=
\frac{s}{\pi^2}\int_0^\infty I_0^{1/2}(s,b,\tilde{b},x_F)J_0(\tilde{b}p_\perp)\tilde{b}d\tilde{b}.
\end{equation}

Since in the model constituent quarks are considered to form a $SU(6)$ wave function,
$I_0=I_0^{U\to S}$.
The function $I_0^{U\to S}(s,b,x_F)$ according to quasi-independent nature
of constituent quark-scattering
can be represented then as a product
\begin{equation}
 I_0^{U\to S}(s,b,x_F)= \left[\prod^{N-1}_{Q=1} \langle f_Q(s,b)\rangle\right]\langle
f_{U\to S}(s,b,x_F)\rangle,
\end{equation}
 where $N$ is the total number of quarks in the colliding
hadrons.

In the model the $b$--dependencies of the amplitudes $\langle f_{Q}
\rangle $ and $\langle f_{U\to S} \rangle $
are related to the strong
formfactor of the constituent quark and transitional spin-flip formfactor
 respectively.
 The strong interaction radius of constituent
quark is determined by its mass. We suppose that the corresponding radius of transitional formfactor
 is determined by the average mass $\tilde{m}_Q=(m_U+m_S)/2$ and factor $\kappa<1$ (which takes
into account reduction of the radius due to spin flip)
$r^{flip}_{U\to S} = \kappa\zeta /\tilde{m}_{{Q}}$ and the corresponding function $f_{U\to S}(s,b,x_F)$ has the form
\begin{equation}\label{ftr}
f_{U\to S}(s,b,x_F)=g_{flip}(x_F)\exp\left(-\frac{\tilde{m}_Q}{\kappa\zeta}b\right)
\end{equation}

 The expression for
$I_0(s,b,x_F)$ can be rewritten then in the following form:
\begin{equation}\label{iol}
I_0(s,b,x_F) =\frac{\bar{g}(x_F)}{g_Q(s)}
U(s,b)\exp[-\Delta m_Q b/\zeta ],
\end{equation}
where the mass difference $\Delta m_Q\equiv\tilde{m}_Q/\kappa-m_Q$ and $\bar{g}(x_F)$ is the function whose
dependence  on Feynman $x_F$ in the model is not fixed.

Now we can consider $p_\perp$- and $x_F$-dependencies
of the $\Lambda$-hyperon production
cross-section and we start with angular distribution\footnote{One should remember that
all formulas and figures below are valid for the fragmentation region only, i.e. for $x_F>0.4$}.
The corresponding amplitude
$F(s,p_\perp,x_F)$ can be calculated analytically. To do so
 we continue the amplitudes
$F(s,\beta, x_F),\,\beta =b^2$, where
\[
F(s,\beta, x_F)=\frac{1}{x_F^2}\frac{I^{1/2}_0(s,\beta,x_F)}{1+U(s,\beta)}
\]
to the complex
 $\beta $--plane and transform the Fourier--Bessel integral over impact
parameter into the integral in the complex $\beta $ -- plane over
the contour $C$ which goes around the positive semiaxis.
Then  for the  amplitude $F(s,p_\perp,x_F)$ the following
representation takes place:
\begin{equation}\label{impl}
F(s,p_\perp,x_F)  =  -\frac{is}{2\pi ^3}\int_C d\beta
F(s,\beta, x_F)K_0(\sqrt{-p^2_\perp\beta/x_F })
\end{equation}
where $K_0(x)$ is the modified Bessel function. The amplitude
 $F(s,\beta, x_F)$ has the poles in the
$\beta $--plane determined by  equation
\begin{equation}\label{poles}
1+U(s,\beta )=0.
\end{equation}
The solutions of this equation can be written as
\begin{equation}
\beta _n(s)=\frac{\zeta ^2}{ M^2}\,\left\{\,\ln g(s)+\,i \pi
n\,\right\},\, n=\pm 1, \pm 3,\ldots \label{polloc}
\end{equation}
 The amplitude $F(s,\beta, x_F)$ besides the poles has a
  branching point at $\beta =0$.

Therefore the  amplitude $F(s,p_\perp,x_F)$  can be represented as
a sum of the poles contribution and the contribution of the cut:
\begin{equation}
F(s,p_\perp,x_F)=F_p(s,p_\perp,x_F)+F_c(s,p_\perp,x_F)\label{sum}
\end{equation}

The poles contribution has an exponential dependence  on $p_\perp$
\begin{equation}
F_p(s,p_\perp, x_F)\simeq isG_p(s,x_F)\sum_{k=1}^\infty  \tau ^k(-p_\perp/x_F)
\varphi_{k}[R(s),p_\perp], \label{lamdif}
\end{equation}
where the parameter $\tau(-p_\perp/x_F)$
\[
\tau(-p_\perp/x_F)=\exp(-\frac{\pi \zeta }{M}\frac{p_\perp}{x_F})
\]
 the function $G_p(s,x_F)$ is
\[
G_p(s,x_F)=\left[\frac{\bar g(x_F)}{x_F^4g_Q(s)}\right ]^{1/2}[g(s)]^{-\frac{\Delta m_Q}{2M}}
\]
and $\varphi_{k}[R(s),p_\perp]$ is the oscillating functions of $p_\perp$, the hadron interaction radius $R(s)$ determines
the period of these oscillations and it has slow energy dependence like $\ln^{1/2}s$.

The cut contributions has power-like dependence on $p_\perp$
\begin{equation}
 F_{c}(s,p_\perp, x_F)\simeq isG_c(s,x_F)(1+\frac{p_\perp^2}{ x^2_F\bar{M}^2})^{-3/2},
\end{equation}
where $\bar M=(M-\Delta m_Q)/2\zeta$ and
the function $G_c(s,x_F)$ has the form
\[
G_p(s,x_F)=\left[\frac{\bar g(x_F)}{x_F^4g_Q(s)}\right ]^{1/2}[g(s)]^{-\frac{1}{2}}
\]
Calculation of poles and cut contributions are similar to the case of
elastic scattering \cite{ech}.

The poles and cut contributions determine the behaviour of
the inclusive cross-section of $\Lambda$ production at moderate and large
values of $p_\perp$ correspondingly, i.e. it will have in
the region of large $p_\perp$ power-like dependence on $p_\perp$:
\begin{equation}\label{dsig}
\frac {d\sigma}{d\xi}\propto G_c^2(s,x_F)(1+\frac{p_\perp^2}{ x^2_F \bar{M}^2})^{-3},
\end{equation}
while at smaller values of $p_\perp$ inclusive cross-section would have the exponential $p_\perp$-dependence:
\begin{equation}\label{dsig1}
\frac {d\sigma}{d\xi}\propto G_p^2(s,x_F)\exp(-\frac{2\pi \zeta }{M}\frac{p_\perp}{x_F}).
\end{equation}
The data for the $\Lambda$-hyperon production are available at
the moderate values of $p_\perp$ and the experimental fits to the
data \cite{data} of the  form \[
A(1-x_F)^n e^{-B(x_F)p_\perp}\]
 just
follow to Eq. (\ref{dsig1}) when relevant parameterization for the
function $\bar g(x_F)$ is chosen. At high values of $p_\perp$ power-like dependence
should take place according to Eq. \ref{dsig}.  In the energy region of $\sqrt{s}\leq 2$ TeV
the functions $G_p$ and $G_c$ have very slow variation with energy due to the numerical values
of parameters \cite{preas}.

\section{Comments on SSA in pion production processes}
SSA is an interesting topic not only in the field of hyperon polarization.
The new experimental
expectations are related to the experiments at RHIC with polarized proton beams and new
experimental data obtained by the STAR collaboration have already demonstrated significant
spin asymmetry in the $\pi^0$--production similar to the one observed
in the fragmentation region at FNAL \cite{e704,rhic}.

We would like to make a brief comment on this subject and to
note that the reverse to the filtering mechanism can be used for the
 explanation of the SSA in pion
production observed at FNAL and recently at RHIC in the fragmentation region.
In the initial state of these reaction the proton is polarized
 and can be represented in the simple SU(6) model as  following:
 \begin{equation}\label{pr}
 p_\uparrow=\frac{5}{3}U_\uparrow+\frac{1}{3}U_\downarrow+\frac{1}{3}D_\uparrow+
 \frac{2}{3}D_\downarrow.
\end{equation}
\begin{figure}[htb]
\begin{center}
  \resizebox{6cm}{!}{\includegraphics*{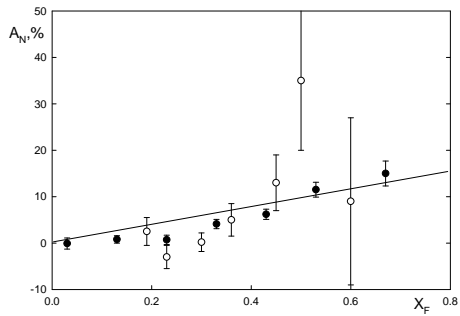}}
\end{center}
\caption{$x_F$-dependence of the asymmetry $A_N$ in  $\pi^0$-production
(filled circles--E704 data \cite{e704}, empty circles-- STAR data \cite{rhic}).}
\label{tsp0}
\end{figure}
The relevant process for $\pi^+$--production is
\[
U_\uparrow\to\pi^+ + D_\downarrow,
\]
which leads to a negative shift in the impact parameter and consequently
to the positive asymmetry $A_N$, while the corresponding process for
the $\pi^-$--production
\[
D_\downarrow\to\pi^- + U_\uparrow .
\]
It leads to the positive shift in impact parameter and respectively
 to the negative asymmetry $A_N$.
Asymmetry $A_N$ in the fragmentation region should have similar to polarization linear $x_F$--dependence which
 is in agreement with the observed FNAL and RHIC experimental data \cite{e704,rhic}
  at $x_F>0.4$.
  As for the neutral
  $\pi^0$--production the combination of $U$ and $D$--quarks with up and down polarization
  makes contributions to cross--sections and asymmetry. On the basis of the simple SU(6)
model we can assume that the $U$--quark with up polarization would contribute mainly
in the fragmentation region. Then the $\pi^0$--production should have positive asymmetry.
The corresponding behavior  and experimental data obtained at FNAL and RHIC are depicted
on Fig. 3. Linear $x_F$-dependence agrees with the experimental data at large $x_F$
(fragmentation region, $x_F>0.4$).

The expression for the unpolarized inclusive cross-section of the pion production in the
fragmentation region can be obtained in a similar way to the corresponding cross-section
of $\Lambda$ production. It should be noted that Eq. (\ref{dsig}) leads to $p_\perp^{-6}$
dependence at large $p_\perp$ and  is valid also
for the pion production. It is in  agreement
with $p_\perp^{-N}$ (with the exponent $N=6.2\pm 0.6$) dependence of
the inclusive cross-section of $\pi^0$-production observed
 in forward region at large $p_\perp$ at RHIC \cite{star}. The exponent $N$ does not depend
 on $x_F$ and choosing relevant function $\bar g(x_F)$ the $(1-x_F)^{5.1\pm 0.6}$ dependence
 of experimental data can be reproduced.

\section*{Conclusion}
The proposed mechanism deals with effective degrees of freedom and takes into
account collective aspects of QCD dynamics. Together with unitarity, which is an essential
ingredient of this approach, it allows  to
 obtained results for polarization dependence on kinematical variables
 in  agreement with the  experimental  behavior
of $\Lambda$-hyperon polarization, i.e.
 linear dependence on $x_F $ and
flat dependence
on $p_\perp$ at large $p_\perp$
in the fragmentation region are reproduced.
Those dependencies together with the energy independent
behavior of polarization at large transverse
momenta are the straightforward consequences of this model.

We discuss here  particle production in the fragmentation region.
In the central region where correlations
 between impact parameter of the initial and impact parameters of the final particles
 being weakened, the polarization cannot be generated due to chiral quark filtering
 mechanism.
Moreover, it is  clear that since antiquarks are produced through spin-zero Goldstone bosons
we should expect $P_{\bar\Lambda}\simeq 0$.
The chiral quark filtering is also relatively suppressed when compared to direct elastic
 scattering of quarks in effective   field and therefore
   should not play a role in the reaction $pp\to pX$ in the fragmentation
 region, i.e. protons should be produced unpolarized. These features take place
 in the experimental data set.

The
application
of this mechanism to description of polarization of other hyperons is more complicated
problem,
since they
could have two or three strange quarks and spins of  $U$ and $D$
quarks can also make contributions into their polarizations.

Finally, it was shown that the mechanism reversed to chiral
quark filtering can provide description of the SSA in $\pi^0$ production
measured at FNAL and recently at RHIC in the fragmentation region and it leads to the
energy independence of the  asymmetry.

\end{document}